# A new paradox in superluminal signaling

Moses Fayngold
Department of Physics, New Jersey Institute of Technology, Newark, NJ 07102

A new paradox in superluminal signaling is presented. In contrast to the Tolman paradox with tachyon exchange between two parties, the new paradox appears already in a one-way signaling, even without creating the time loop. As shown in the article, a one-way superluminal signaling would, by swapping cause and effect, violate the factual invariance of some related physical processes. This produces a universal ban on any kind of superluminal signals, which is much stronger than the Tolman paradox. Even though relativity embraces superluminal motions as such, thus making the world symmetric with respect to the invariant speed barrier, the ineptness of all such motions for superluminal signaling makes the symmetry incomplete.

## 1. Introduction

As is well known, superluminal signaling (SS) as described in the wave theory is forbidden by the Sommerfeld-Brillouin theorem, according to which an edge necessary for a wave-form to transfer information cannot exceed the invariant speed $c$ [1, 2]. Generally, SS would violate causality in some processes of signal exchange between two objects (the Tolman paradox [3]). This work describes a new paradox which appears already in one-way SS.

We will identify the SS-carriers with hypothetical tachyons [4-10] and consider them classically, as was done in [3]. In order for tachyons to be observable, they must interact with ordinary matter through emission, scattering and absorption. We consider a tachyon interaction with two objects A and B. Hereafter, we denote an *object* by a straight capital symbol, and an *event* in object's history – by the corresponding *italics* capital.

According to the Lorentz transformations, for any positive-energy tachyon connecting events $A$ and $B$ in an inertial reference frame (RF) K, one can find another inertial frame $\mathrm{K}'$ where the tachyon's energy is negative and the ordering of $A$ and $B$ is time-reversed. But according to the reinterpretation principle (RIP), the negative-energy tachyon moving backward in time from $A$ to $B$ in $\mathrm{K}'$, must be actually observed as positive-energy tachyon moving forward in time from $B$ to $A$ [4-7]. In all such cases, cause and effect exchange their roles: $A$ is the cause of $B$ in K, but is its effect in $\mathrm{K}'$. This saves the causal ordering postulate (COP) (cause always precedes the effect) at the cost of losing Lorentz-invariance of its constituents.

Impossibility of SS was proven for a complete cycle: sending a signal to a remote receiver and obtaining the response. Under certain conditions, the response comes to the sender before the emission of the original signal, thus creating the Tolman paradox. Another version of the paradox involves 4 observers moving along the respective sides of a square [9]. The simplest way to avoid paradoxes while conserving tachyons was to forbid the "tachyon exchange". As Nick Herbert put it, "*Some physicists, noting that all time-travel paradoxes arise from returning to a location before you left it, decided to eliminate the paradox by refusing to issue round-trip tickets to tachyons*" [11].

The attitude towards a *one-way* SS is more tolerant, which is somewhat surprising in view of the well-known result [1]. The major (but not unanimous) consensus is that one-way SS-s, while looking weird for some observers, might be compatible with both – Special Relativity (SR) and causality as far as it is not used for signaling back.

However, despite many attempts, the superluminal particles have not been discovered (for the recent case, see, e.g., [12-14]). Reported observations of SS in quantum tunneling [15-18] are just another case of superluminal group velocity, which cannot be harnessed for SS [19-22]. One specific claim of SS allegedly observed in frustrated internal reflection [17, 18] was shown to be based on the wrong interpretation of the tunneling process [23]. In terms of daily observations, stability of the atomic ground states is compelling evidence against tachyons. As was shown in [24, 25], tachyon interactions with atoms must make such states unstable – sufficiently fast moving atoms could spontaneously get excited to higher states by *emitting* tachyons at the cost of atoms' kinetic energy. Such a process has never been observed, which may well be due to nonexistence of tachyons.

With all that, there was, to our knowledge, no rigorous theoretical proof (apart from result [1] in the wave theory) of the general impossibility of SS. Such a proof is the goal of this work. We will first review briefly some relevant properties of a single tachyon carrying 1 bit of information (Sec. 2) and then of succession of *n* tachyons transmitting a superluminal message (Sec. 3). Then it will be shown in Sec. 4 that already a one-way SS leads to contradictions, so SS must be forbidden in principle.

## 2. Basic features of superluminal signaling

Here we will revise some features of tachyon dynamics. One of them is the Cerenkov radiation (CR) which must generally be emitted even by a neutral tachyon [26]. On the quantum level, CR would cause tachyon's Brownian motion around its unperturbed trajectory [27]. In the classical limit, the CR is expected to be axially-symmetric, but such symmetry holds only in a special subset of RF. Generally, the tachyon's form-factor and CR are *asymmetric*, and its trajectory is curved. This makes the tachyons unreliable for signaling along a chosen direction [27-29].

The energy loss due to CR leads to $E \to 0$ and then to $E$ getting negative. According to RIP, the tachyon at this point must experience head-on collision with its anti-self coming from the opposite direction [24, 25, 27]. This point determines a restriction on distances available for one-way SS with given initial energy. But this falls short of an absolute ban, since we can assume the distance between A and B within the allowed range. And for certain tachyon types the probability of re-absorption of CR can be close to that of its emission, so the rate of the energy loss to CR is very low [30]. All these factors allow us to neglect the effects of CR as had been done in the Tolman paradox.

Suppose that object A is initially stationary in K, while B is stationary in $K'$, but both objects can be observed from either frame. Let $\tilde{v}$ stand for the tachyon speed in K, and $V$ for the relative velocity between $K'$ and K. Let A emit a tachyon (event *A*), which is subsequently absorbed by B (event *B*) (Fig.1).

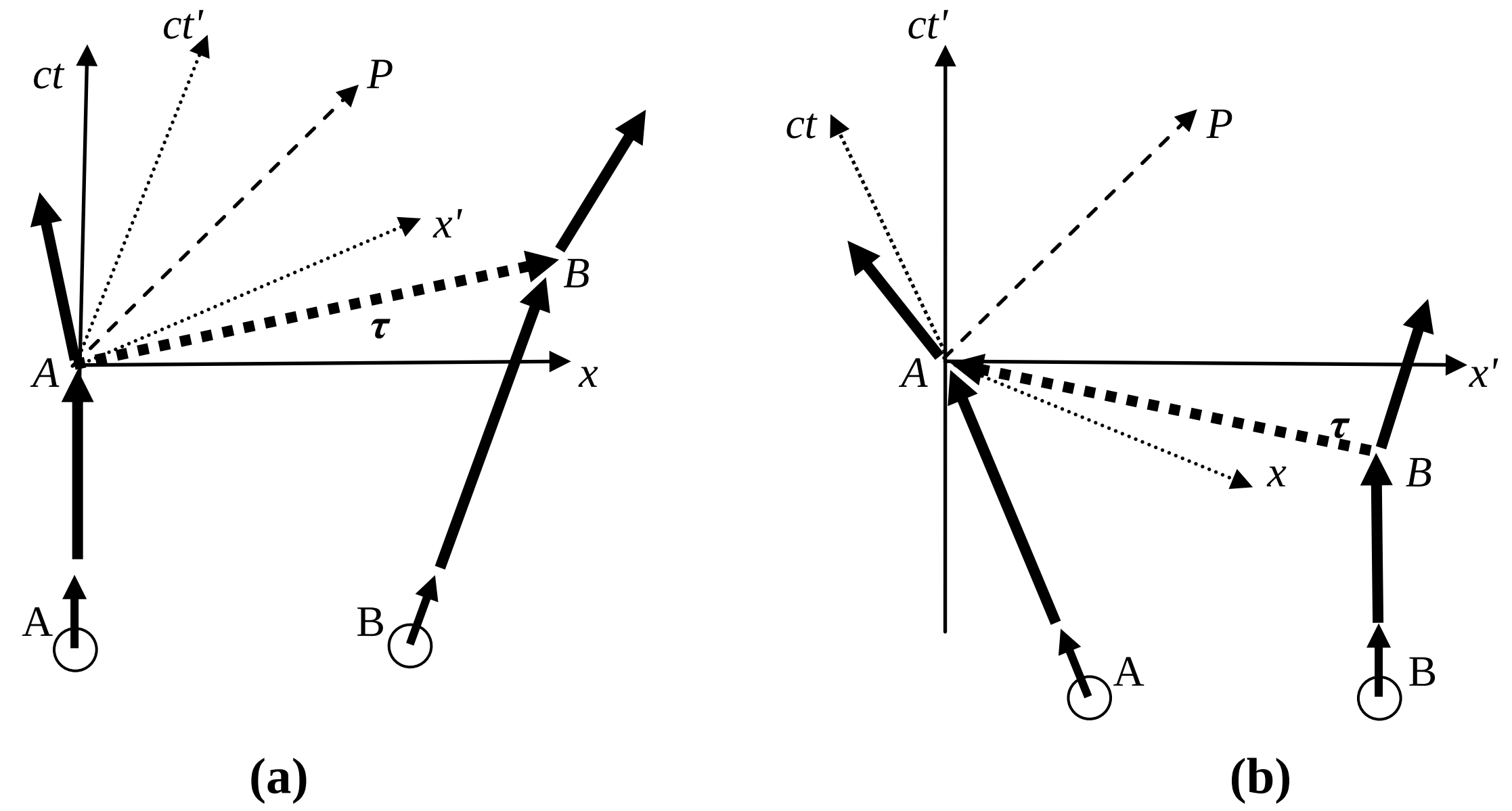

**Fig. 1**

Space-time diagrams of objects A and B connected by a tachyon signal

(a) – as observed in K; (b) – as observed in $K'$.

The solid lines are the world-lines of the objects. They are broken (recoil) at points *A* and *B* due to interaction with tachyon $\tau$ (world line *AB* in K and *BA* in $K'$). AP is the world line of a photon passing through the origin of the respective RF.

Since the tachyon's world-line between *A* and *B* is space-like, the time ordering of events (*A*, *B*) in $K'$ may be opposite to that in K. This happens when [3, 10, 24, 25]

$$V\tilde{v} > c^2 \qquad (1)$$

Vectors $\tilde{\mathbf{v}}$ and $\mathbf{V}$ are assumed collinear and A and B are receding from each other.

The fact that signal transfer from A to B in K is, under condition (1), observed as going from B to A in $K'$ can be expressed symbolically as:

$$A \xrightarrow[\tau,\ \text{in K}]{} B \quad \text{and} \quad A \xleftarrow[\tau',\ \text{in K}']{} B \qquad (2)$$

Both expressions in (2) describe *the same* one-way SS *with a single tachyon* $\tau$ observed from two different RF. The recorded change of direction has nothing to do with familiar effect observed when outrunning a fast-moving object. One cannot outrun a tachyon. Expressions (2) are just mathematical formulation of RIP [4, 6, 24, 25].

The process (2) is quite different from *a round-trip communication* $A \rightleftarrows B$ involving *two* tachyons – the primary $\tau_1$ emitted from A to B, and the secondary $\tau_2$ – the response from B sent back to A. This tachyon exchange can be recorded as a *pair of one-way* communications in *either* RF. If the speed of either tachyon satisfies (1), then according to RIP, *both* tachyons move from A to B in K, and from B to A in $\mathrm{K}'$. Symbolically:

$$\mathrm{A} \underset{\tau_2}{\overset{\tau_1}{\rightrightarrows}} \mathrm{B} \ \text{in K} \qquad \text{and} \qquad \mathrm{A} \underset{\tau_2}{\overset{\tau_1}{\leftleftarrows}} \mathrm{B} \ \text{in K'} \qquad (3)$$

(Of course, their *actual world lines* in spacetime are generally not parallel). If, in addition, B responds sufficiently fast, then $\tau_2$ arrives at A (or is emitted by A, depending on RF) *before* the emission of $\tau_1$. This creates the Tolman paradox (Fig. 2).

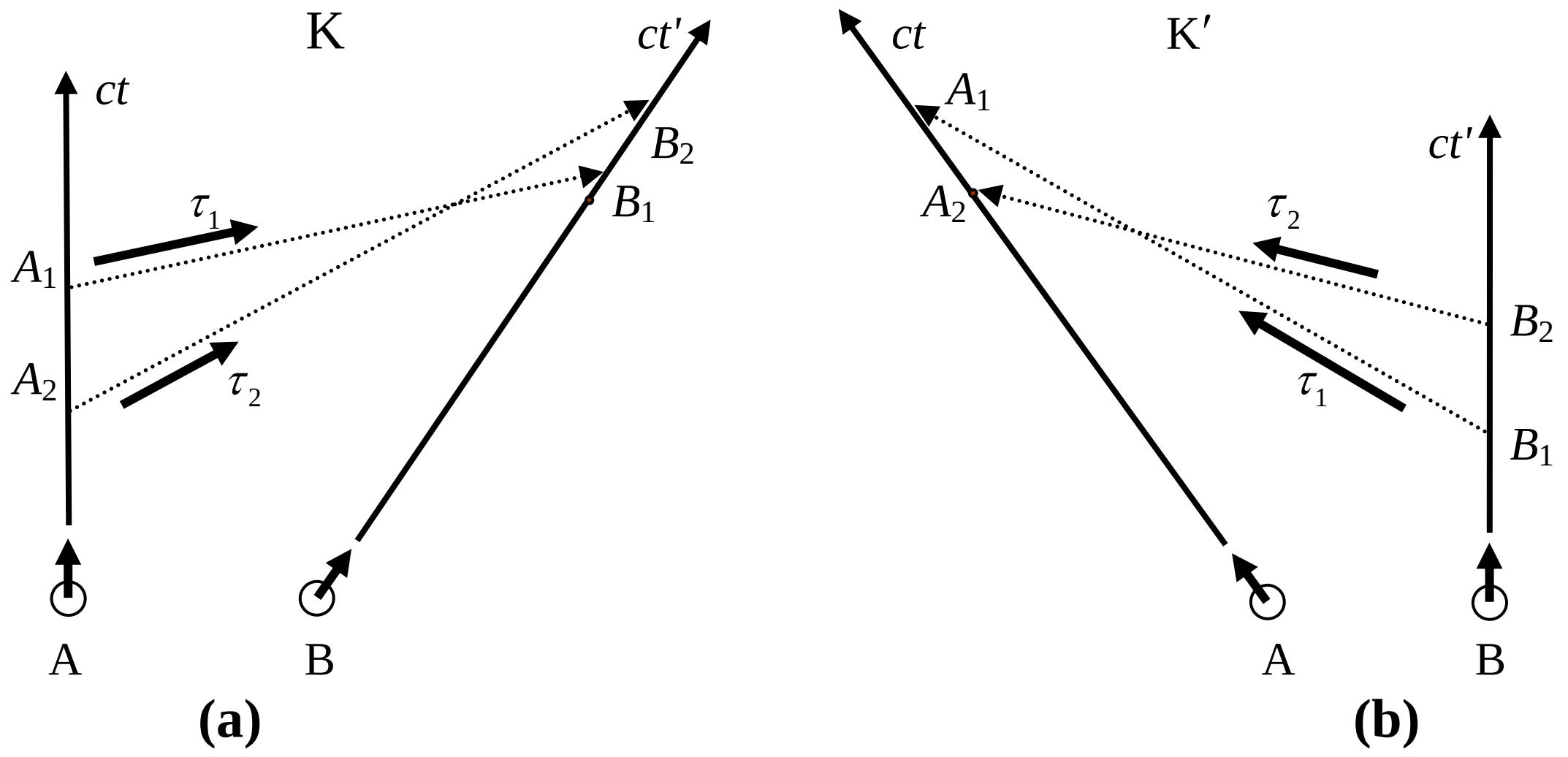

**Fig. 2**

Tachyon exchange between A and B, forming the time loop (the Tolman paradox).

(a) – The process as seen from the rest frame K of object A. The secondary tachyon $\tau_2$ (response from receding B) is observed as spontaneously emitted from A (event $A_2$) *before* the emission of primary $\tau_1$ (event $A_1$ ). This earlier emission might preclude the emission of $\tau_1$.

(b) – The same process as recorded in the rest frame $\mathrm{K}'$ of B. The primary $\tau_1$ from A is observed as emitted from B (event $B_1$) *before* its emission from A. The response $\tau_2$ (event $B_2$) outruns $\tau_1$ and arrives at A before the absorption/emission of $\tau_1$ – the same result as observed in K. The recoil effects are assumed here to be negligible.

Getting back to one-way SS, let us introduce observer Alice in K and Bob in $\mathrm{K}'$. Suppose A, B are the atoms with the ground and excited states $|0\rangle_{\mathrm{A}}$, $|1\rangle_{\mathrm{A}}$ and $|0\rangle_{\mathrm{B}}$, $|1\rangle_{\mathrm{B}}$, respectively. While Alice sees tachyon emission by A followed by its absorption at B, Bob records tachyon emission from B followed by its absorption at A. *For a one-way space-like world line,* this interpretation is consistent with known principles [4-11].

But the events at the *end points* of path *AB* in Fig. 1 may look strange. Atom A as observed in K emits the tachyon in the optical-type transition $|1\rangle_{\mathrm{A}} \to |0\rangle_{\mathrm{A}}$. Atom B absorbs this tachyon, but contrary to our intuition suggesting its "jumping up" from $|0\rangle_B$ to $|1\rangle_B$, the actual transition will be $|1\rangle_B \to |0\rangle_B$. This seems exotic to Alice but is allowed in tachyon-tardyon interactions [10, 25]. The energy released in this transition plus energy of the absorbed tachyon both add to kinetic energy of receding B. Since atom's "personal" history is frame-independent, the same transitions will be recorded in $K'$. But there, $|1\rangle_B \to |0\rangle_B$ comes with tachyon *emission* which looks natural to Bob. By the same token, while $|1\rangle_A \to |0\rangle_A$ is accompanied by tachyon emission in K, Bob records tachyon absorption, with the same explanation as absorption at B seen by Alice. Two emissions – one from A as observed in K, and the other from B as observed in $K'$, are totally equivalent, and the same is true for absorptions. This reflects the symmetry in the initial conditions – each atom is initially at rest in its respective RF.

The situation may be different in the absence of such symmetry, e.g., if *both* atoms were stationary say, in K while Bob remains in $K'$ [24]. Then, the tachyon absorption by B would require transition $|0\rangle_B \to |1\rangle_B$. This seems miraculous to Bob since he observes the tachyon *emission* from B. But again, such process is not forbidden by conservation laws for tachyons. If B is part of K, it is moving in $K'$, and part of its kinetic energy may, as mentioned in Sec. 1, convert into the energies of its internal excitation and the emitted tachyon [10, 24, 25]. In other words, even an atom in its ground state could be generally unstable with respect to tachyon emission. Such instability can therefore be considered as the cause of emission, and accordingly the process at B is causal. And it would be OK with Alice, since in her frame this process would be observed as tachyon *absorption*.

Thus, the above analysis leads to a theoretical prediction:

*If tachyons do exist and interact with ordinary matter, then even the objects' ground states must be unstable. A sufficiently fast-moving atom in a state $|0\rangle$ could undergo spontaneous transition $|0\rangle \to |1\rangle$ to an excited level and emit a tachyon in the forward direction – all at the cost of its kinetic energy. This would be observed as a tachyon absorption in the atom's rest frame.* Alternatively, an atom could *absorb* tachyon in transition $|1\rangle \to |0\rangle$. Then the released internal energy and the energy of absorbed tachyon would both go to increase of atom's kinetic energy.

Of course, the considered frame-dependent instability of atomic state $|0\rangle$ is purely kinematic effect, so it falls short of the innate instability of excited states. But with RIP accepted, all states can be considered on the same footing in terms of causality. In this respect, both events *A* and *B* are causally equivalent. In K, event *A* initiates the transaction described by the left Eq. (2). In $K'$, this transaction (the right of Eq. (2)) is time-reversed and is initiated by *B*.

This reversibility must be also manifest in the evolution of the tachyon itself during its travel between A and B. To show it in a dramatic way, let us represent each tachyon as a virtual alien whose lifetime equals the corresponding travel time between A and B. Suppose that in K, the alien evolves from a newborn at the emission time $t_A$ to an old one during the travel and dies at the moment $t_B$ of arrival at B. However, for a $K'$-observer, *the same* alien emerges as an old one from B, gets younger on the way to A, and arrives at A as a baby to be absorbed.

In a more realistic case, a tachyon may be represented by an unstable particle in the Gamow state (assuming such states possible for tachyons). Then we can, by analogy with regular bits, denote the tachyon's higher and lower inner energy states as $|1\rangle_\tau$ and $|0\rangle_\tau$ respectively. The tachyon decay (with a *tardyon* emission) recorded as $|1\rangle_\tau \to |0\rangle_\tau$ in K, could be observed as $|1\rangle_\tau \leftarrow |0\rangle_\tau$ in $K'$ even though the emitted tardyon would still be receding from it. That would be another exotic feature of SS.

Some arguments invoke entropy as a measure of time:

"*Let us replace the tachyon with something that has a clock, e.g., an aging spaceship with paint peeling off... Then one can use entropy as a clock. One can no longer have observers decide to re-interpret the time-ordering; the direction of time is given by the ordering that shows the space-ship aging. So the RIP must be invalid.*" [31]

But references to the entropy of a superluminal object as an effective clock do not seem logically conclusive, because even if we could read tachyonic clocks, such readings would not express *our* time which is determined by *tardyonic* clocks. Our history is irreversible regardless of RF. But a tachyonic spaceship with its inside dilapidating from freshly painted to peeled off in K, evolves from peeled off to freshly painted in $K'$. This is a purely kinematic effect for SS. The time direction is not determined by system's entropy [32]. And the respective records from both – K and $K'$ – are equally legitimate. So neither of the considered examples can rebut RIP, and we could as well have started this section from considering first Bob's records in $K'$.

### 3. The information flow in structured superluminal signaling

Consider now a *structured* SS, containing *n* bits, numbered as $1, 2, ..., j, ..., n$. We assume that each bit is a tachyon in one of the two possible states. The emission of each bit by A in K is a separate event characterized by the corresponding 4-vector $s_A^j \equiv (t_A^j, \mathbf{r}_A^j)$. Here $t_A^j$, $\mathbf{r}_A^j$ are, respectively, the moment of emission of *j*-th bit, and its position at this moment in K. For A being stationary in K, all $\mathbf{r}_A^j = \mathbf{r}_A = const$. Similar notations, $s_B^j \equiv (t_B^j, \mathbf{r}_B^j)$, will be used for absorption events at B, and the same in frame $K'$, with $t$, $\mathbf{r}$ primed. The whole succession of bits forms an extended message emitted by A during the time interval $\Delta t_A \equiv t_A^n - t_A^1$ around the central moment $t_A$, and absorbed by B within the time interval $\Delta t_B \equiv t_B^n - t_B^1$ around $t_B$.

We must distinguish between the time ordering of events $\mathbf{s}_A^j, \mathbf{s}_B^j$ for a *fixed* bit No. *j* (Sec. 2), and time ordering of *emissions/absorptions of bits* within $\Delta t_A$ (or $\Delta t_B$). The former is frame-dependent, whereas the latter is Lorentz-invariant. Suppose, A sends to B a superluminal message "*Meet the tachyons!*". Each pair of events "emission-absorption" $\left(\mathbf{s}_A^j, \mathbf{s}_B^j\right)$ *of the same bit* forms a space-like interval, therefore succession of these events is time-reversed in $K'$ under condition (1): each bit is first *emitted* from B and later absorbed by A, as recorded in $K'$. On the other hand, emission/absorption of two consecutive bits, say, $\left(\mathbf{s}_A^j, \mathbf{s}_A^{j+1}\right)/\left(\mathbf{s}_B^j, \mathbf{s}_B^{j+1}\right)$, forms a time-like interval, so the corresponding Lorentz transformation does not affect their ordering. The tracking records of each separate bit in K and $\mathrm{K}'$ are time reversed of one another, but the structure of the message is the same. In $\mathrm{K}'$, the tachyons are departing from B, but their succession still reads "*Meet the tachyons!*". A superluminal message is robust and its structure is invariant even when departure-arrival ordering and thereby the life story of each separate bit (tachyon) is time-reversible under an appropriate Lorentz transformation. The succession of bits, e.g., "first bit" and "the last bit" in the message from A to B remains unchanged when observed as message from B to A in $\mathrm{K}'$. Symbolically, using the above-introduced notations, we have

$$\left.\begin{array}{c} t_A^j < t_B^j \\ \Delta t_A > 0,\ \Delta t_B > 0 \end{array}\right\} \text{in K}, \qquad \text{and} \qquad \left.\begin{array}{c} t_A'^j > t_B'^j \\ \Delta t_A' > 0,\ \Delta t_B' > 0 \end{array}\right\} \text{in K}' \qquad (4)$$

Thus, the amount and content of superluminal information, being determined by the number and succession of bits, are Lorentz-invariant, while the direction of its flow may be frame-dependent. For collinear motion of bits and for used RF, this direction is, under condition (1), determined by expressions (2).

A widely spread misconception is confusing information flow with information contents, as seen from the following example. Suppose that a message is the text of "Hamlet" sent as a modulated tachyonic beam by Shakespeare himself from A to Bacon at B [7, 24]. But under condition (1), it would be recorded in $\mathrm{K}'$ as flowing from B to A, even though its emission from B might be centuries of $\mathrm{K}'$-time before Shakespeare's birth. This seems weird in view of the fact that Shakespeare is considered to be the indisputable author of "Hamlet". The latter may seem a compelling argument for the following conclusion in [7] (cited in italics): "…*no amount of reinterpretation will make Bacon the author of Hamlet. It is Shakespeare, not Bacon, who exercises control over the content of the message. For any tachyon trajectory the time ordering of the end points is relative...But the direction of information transfer is necessarily a relativistic invariant*."

All statements here preceding the last one are correct. But they are not sufficient to justify the last statement in a world with tachyons. The direction of information transfer is determined entirely by the motion of signal carriers, not by the physical conditions at the end points of their trajectory. The fact that text of Hamlet has a special meaning for some humans cannot be used as a physical argument for "supremacy" of its prehistory at A over conditions at B. Accordingly it is not an argument in favor of universal flow of information from A to B in all RF. The tachyon emissions from B as observed in $\mathrm{K}'$ also require certain preconditions. For instance, if B is part of $\mathrm{K}'$ receding from A, then its atoms must be, as discussed in Sec. 2, uniquely excited in order to properly *absorb* the message as viewed from A and thus record the arriving signal, just as A-atoms must be uniquely excited to emit the corresponding tachyonic bits. This unique tuning also accounts for the emission of a structured text from B and its absorption at A, as recorded in $\mathrm{K}'$. If A and B are both stationary in K, then the respective B-atoms must be all in the ground states to properly absorb the sequence of arriving tachyons in K, or properly emit them from the viewpoint of $\mathrm{K}'$. In either case there is a specific prehistory leading to emissions from B as observed in $K'$. But direction of information flow is determined by direction of information carriers, and the latter is determined by RIP.

There are still some important questions to be answered, e.g., why all the bits emitted from B are all directed towards A. One plausible explanation might be that it is a very unlikely but physically possible fluctuation. But even that leaves the issue debatable. For instance, getting back to Alice-Bob communications, Alice could in each trial decide – to shoot or not to shoot at B from her tachyon gun, depending on the outcome of a coin toss. In all cases of not shooting, Bob would not observe any emissions from B. Thus, it turns out that for Bob all such emissions or absence thereof are rigidly correlated with the *future* conditions at A. This is already violation of causality in $K'$, but it is perfectly OK from Alice's perspective. Thus, the question of which object, A or B, is the actual source of the corresponding tachyonic signal, may still cause disagreement.

To simplify the following discussion, we reduce the process to a single and stable tachyon connecting two single atoms A and B, stationary in K and $K'$, respectively, with Alice and Bob only registering the events. This avoids the possible controversies of the type "who sent what", and we could consider a tachyon emission in either frame as a spontaneous event in that frame. Some results may still challenge our "subluminal" intuition, but they do not seem by themselves to contradict basic principles. If the A-atom emits a tachyon with $\tilde{v} > c^2/V$ towards B, Bob observes the B-atom emitting this tachyon toward A. In the absence of emission from A there are no emissions from B, and vice-versa.

So, summarizing this part, we can say that a continuous one-way SS may, on the face of it, appear legitimate for all observers, at least on the atomic level. The records obtained in K and $K'$ have certain symmetry with respect to each other in terms of both – laws of nature and initial conditions. This may explain the apparently peaceful coexistence of relativistic causality and the concept of signal-carrying tachyons.

## 4. Insertion of an intermediate absorber

This, however, changes dramatically if we consider tachyon interaction with some other object between A and B. Let us introduce a third party, Celia, whose world line C runs between A and B (Fig. 3).

On the verge of crossing the tachyon world line *AB* Celia makes a decision – to put or not to put an absorber C in the tachyon path. The C may be a plate opaque for tachyons within a broad energy range. Now the conditions at A and B are equivalent, and event *C* happens when the tachyon is *indisputably on its way* from A to B or from B to A, depending on the observer. Denot the K-moments of tachyon emission at A, its passing by C, and subsequent absorption at B as $t_A$, $t_C$, and $t_B$, respectively. These events are observed in $K'$ (in the reversed chronological order) as emission at B, passing by C, and absorption at A. Their $K'$-moments are, respectively, $t'_B$, $t'_C$, and $t'_A$. We have

$$\left.\begin{array}{c} t_A < t_C < t_B \\ \text{in K} \end{array}\right\} \quad \text{and} \quad \left.\begin{array}{c} t'_A > t'_C > t'_B \\ \text{in K}' \end{array}\right\} \tag{5}$$

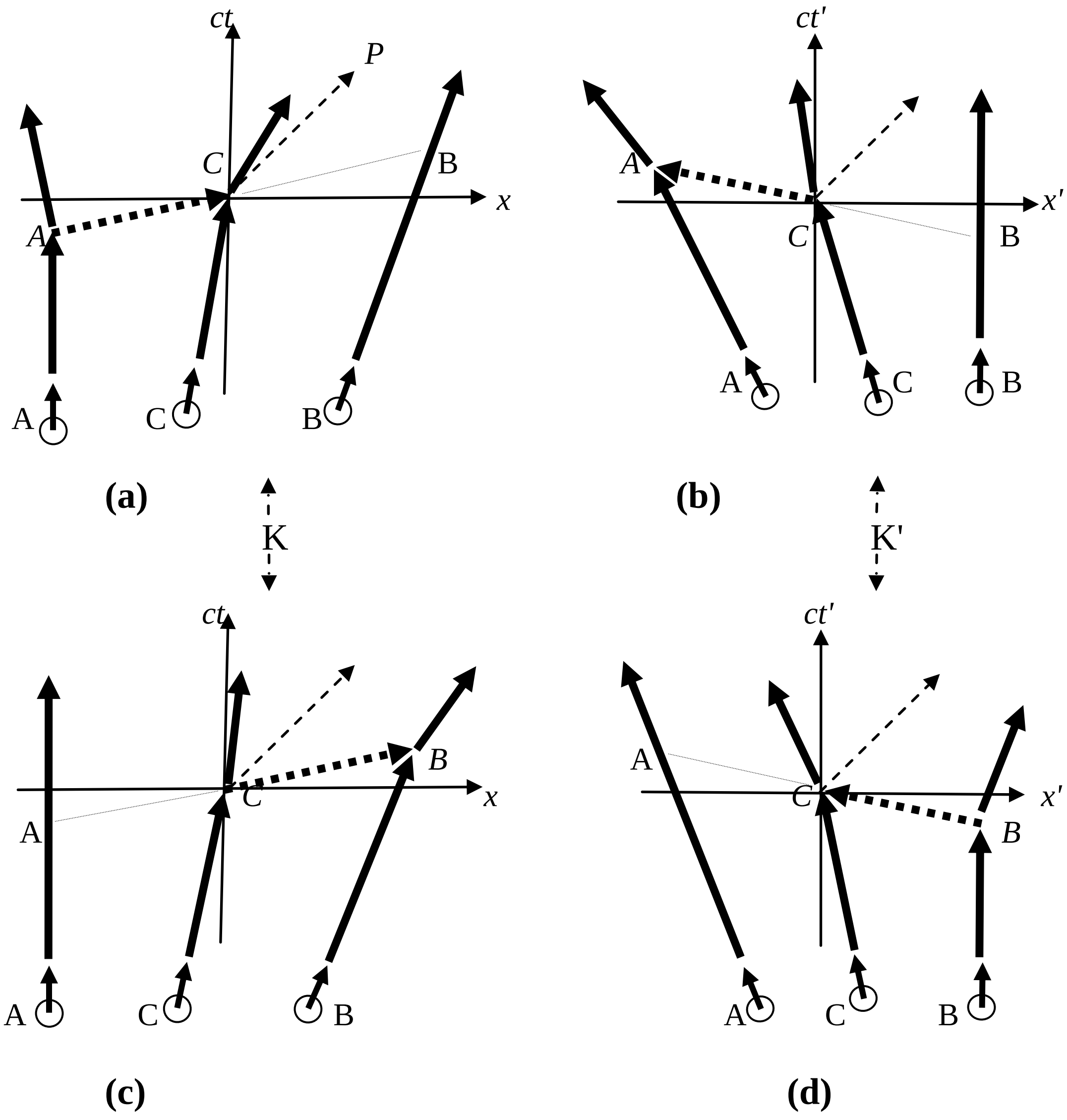

**Fig. 3**

Objects A and B connected by a single tachyon. The tachyon's interception by an intermediate absorber C produces paradoxical situation: neither of possible outcomes is compatible with known physical principles.
(a), (b) – Alice's expectations of the outcome (as seen in frames K and K′, respectively)
(c), (d) – Bob's expectations of the outcome (also as recorded in K and K′).
Their predictions are mutually exclusive.

Suppose Celia decides to act, and executes her decision. This immediately creates a dilemma – which part of the tachyon world line will be blocked by C.

In K, the B-bound tachyon from A is intercepted by C ( $A \to C$, Fig.3a), so the part *C*B does not materialize. Also, both – A and C – are kicked away from each other. The pair of objects interacting with tachyon is (A, C), while B remains idle. Due to factual invariance of events, Alice expects the same pair to be recorded by Bob, only in the reversed order, as (C, A). In other words, Bob is expected to record the tachyon emission from C and its subsequent absorption by A, with B inactive ( $A \leftarrow C$ , Fig.3b).

But it is not so from Bob's perspective! According to (5), Celia's action as observed in $K'$ happens well *after* event *B*, so it must block segment *C*A, but it cannot affect the preceding part *BC*.

To take a closer look, suppose that right before Celia's action she receives a telephone call that distracts her from it. In all those trials when Celia restrains from action, Bob observes the tachyon traversing the whole path $A \leftarrow B$. The same must be the case now. But in order to make it, the tachyon *must* have first progressed along the way $C \leftarrow B$ by moment $t'_C$. The part (*B*, *C*) of the tachyon's trajectory is as real for Bob as is part (*A*, *C*) for Alice. Therefore, when Celia does insert C at $t' = t'_C$, it follows from Alice's version that it must cancel *already pre-existing* part ( $C \leftarrow B$ ) of tachyon's history in $K'$, including the atomic transition $|1\rangle_B \to |0\rangle_B$ with tachyon emission at $t' = t'_B$. Canceling *already evolving* process by a future event is an outrageous violation of causality. Thus, Alice's view upholding supremacy of event *A*, leads to contradiction in $K'$.

But Bob's view is the same blind venue as the Alice's one. Bob insists that Celia's action at *C* must block the subsequent part (*C*, A) of the process, so Alice can observe only the prior stage $C \leftarrow B$, and in the reversed order, as $C \to B$. The pair of objects interacting with tachyon and accordingly kicked away from each other is (B, C), whereas A remains idle. But that would mean the same clash with causality, now from the viewpoint of K.

Both scenarios flatly contradict one another, and since they both describe the factual features of the same process rather than its numerical characteristics which might be relative, this is a logical contradiction. Accepting the view of either partner leads to causality violation in the other's RF. And in contrast with the twin paradox, which is easily resolved by noticing non-equivalence of the two used RF, there is no resolution here since both used RF are totally equivalent.

## 5. Summary

The simple thought experiment in the previous section may have 4 possible outcomes:

(a) Both observers record the pair of events (*A*, *C*) (albeit in the opposite ordering)
(b) Both record the pair (*B*, *C*), also in opposite ordering
(c) Alice records (*A*, *C*), whereas Bob records (*B*, *C*)
(d) Alice records (*C*, *B*), whereas Bob records (*C*, *A*)

The outcome (a) violates causality in $K'$ – event *C* cancels *preceding* process (*B*, *C*). Outcome (b) violates causality in K: event *C* cancels preceding process (*A*, *C*). Both outcomes (c) and (d) violate the factual invariance, according to which all observers must record *the same* pair of events.

All 4 possibilities contradict basic principles and therefore must be discarded. Thus, SS leads to paradoxes already in one-way communications. Paraphrasing Nick Herbert, if we adhere to basic principles, *we must refuse to issue even one-way tickets to SS-carriers*. And this would disqualify tachyons from transmission of superluminal interactions.

The only remaining option is to ban the originally assumed tachyon interactions with ordinary matter, so that a tachyon beam could not be obstructed by C. But that would cancel interactions with A and B as well, and thus eliminate any possibility of SS.

The presented analysis has been criticized [33] after publication of the previous version of this work, but the criticism was shown to be unsubstantiated [34].

*Conclusion*: Even though superluminal motions as such do exist and are consistent with relativity, none of them can be harnessed for SS. The considered paradox provides a universal ban on SS, independent of the Sommerfeld- Brillouin theorem. It is much stronger than the Tolman paradox which appears only in the round-trip communications.

### *Acknowledgements*

I am grateful to Keith Fredericks for pointing at some aspects of Cerenkov radiation by tachyons, Art Hobson for valuable comments, and Anwar Shiekh for thorough discussion and constructive criticism.